\documentclass{camera}
\usepackage{graphicx}  

\begin{document}

%
\title{Isoscaling and the symmetry energy in spectator fragmentation}

%
\author{ W.~Trautmann$^{1)}$, A.S.~Botvina$^{1,2)}$, J.~Brzychczyk$^{3)}$,\\
A.~Le F{\`e}vre$^{1)}$, P.~Paw{\l}owski$^{4)}$, 
C.~Sfienti$^{1,5)}$\\
and the ALADIN and INDRA collaborations
}

%
\organization{
$^{1)}$ Gesellschaft f{\"u}r Schwerionenforschung, D-64291 Darmstadt, Germany\\
$^{2)}$ Institute for Nuclear Research, 117312 Moscow, Russia\\
$^{3)}$ M.~Smoluchowski Institute of Physics, Pl-30059 Krak{\'o}w, Poland\\
$^{4)}$ H. Niewodnicza{\'n}ski Institute, Pl-31342 Krak{\'o}w,Poland\\
$^{5)}$ Dipartimento di Fisica and LNS-INFN, I-95126 Catania, Italy
}

\maketitle

\begin{abstract}
Isoscaling and its relation to the symmetry energy in the fragmentation 
of excited residues produced at relativistic energies were studied in 
two experiments conducted at the GSI laboratory. The INDRA multidetector 
has been used to detect and identify light particles and fragments with 
$Z\leq 5$ in collisions of $^{12}$C on $^{112,124}$Sn at incident energies of 300 
and 600 MeV per nucleon. 
Isoscaling is observed, and the deduced parameters decrease with increasing 
centrality. Symmetry term coefficients, deduced within the statistical 
description of isotopic scaling, are near $\gamma =$ 25~MeV for peripheral 
and $\gamma <$ 15~MeV for central collisions. 

In a very recent experiment with the ALADIN spectrometer, the possibility 
of using secondary beams for reaction studies at relativistic energies 
has been explored. Beams of $^{107}$Sn, $^{124}$Sn, $^{124}$La, and $^{197}$Au were
used to investigate the mass and isospin dependence of projectile 
fragmentation at 600 MeV per nucleon. The decrease of the isoscaling 
parameters is confirmed and extended over the full fragmentation regime 
covered in these reactions.
\end{abstract}

%
\section{Introduction}

The symmetry energy and its density dependence 
have received increasing attention in recent years because of 
their importance for nuclear structure and for astrophysics.
Supernova simulations or neutron star models require inputs for the nuclear 
equation of state at extreme values of density and asymmetry 
\cite{lattprak,botv04}. The strength of the symmetry term at about 
normal density
is quite well known from the masses of finite nuclei but its density 
dependence is experimentally very poorly constrained. Theoretical predictions 
are quite consistent for low-density nuclear matter \cite{baldo} but diverge 
considerably at higher density where they are important for astrophysical
phenomena as, e.g., the cooling of neutron stars \cite{bao02}. 
Nuclear reactions offer the possibility to investigate nuclear matter 
at densities
other than normal nuclear density, and an active search for observables 
suited to probe the strength of the symmetry term in reaction experiments is 
presently underway \cite{bao02,greco02,baoan}.

Multifragmentation is generally considered a low-density phenomenon. 
However, the short-range nature of the nuclear forces causes a clustering 
in nuclear systems that have expanded as a result of an initial 
compression or heating in the course of a violent nuclear collision.
Nuclear matter is not homogeneous under such conditions and the strength 
of the symmetry term is thus not easy to predict. 
In statistical multifragmentation models,  
it is assumed that normal-density fragments are distributed 
within an expanded volume. The density is only low on average, and standard 
liquid-drop parameters are used to describe the nascent fragments including 
their isotopic degrees of freedom. In the Copenhagen version of this 
model (SMM),
the symmetry energy term $E_{\rm sym} = \gamma (A-2Z)^2/A$ is used with
coefficients in the range $\gamma =$~23 to 25~MeV \cite{bond95,botv02}.

\begin{figure}[htb]
\centering
\includegraphics[height=10.0cm]{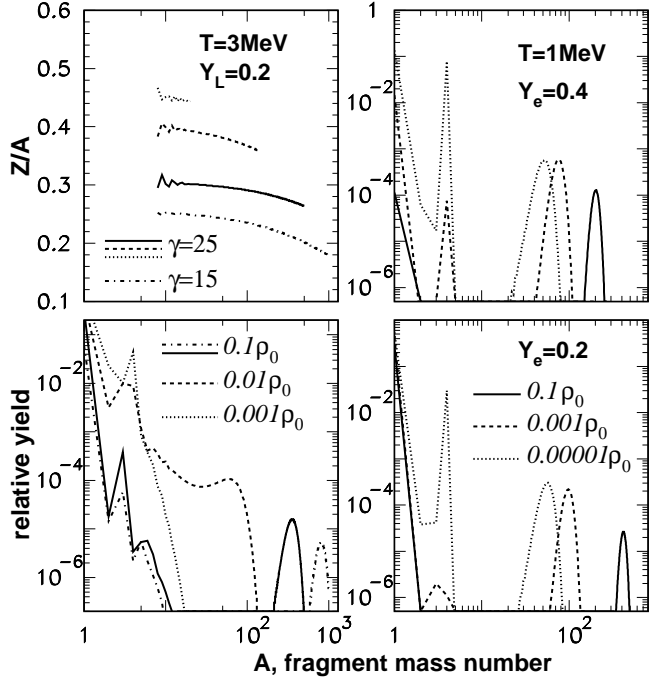}
\caption{
Mean charge-to-mass ratios (left top panel), and mass distributions 
of hot primary fragments (other panels) calculated with the SMM 
generalized for supernova conditions. Results are shown for $T=3$~MeV 
and fixed lepton fraction $Y_{\rm L}=0.2$ (left panels) and $T=1$~MeV and 
fixed electron fractions $Y_{\rm e}=0.4$ and 0.2 (right panels). Baryon 
densities are chosen as indicated. The dash-dotted lines represent 
calculations for density 0.1 $\rho_0$ and a reduced symmetry term coefficient
$\gamma =15$~MeV (from \protect\cite{botv05}).
}
\label{fig01} 
\end{figure}

The knowledge of the strength of the symmetry term under such conditions
is of particular interest because the subnuclear densities and the 
temperatures reached at the freeze-out in fragmentation reactions 
overlap with those expected 
for the explosion stages of core-collapse supernovae \cite{botv05}. 
This similarity permits laboratory studies of the properties of nuclear 
matter in the hot environment similar to the astrophysical situation. 
Nuclear matter in equilibrium at low density is expected to 
consist of nuclei with a wide range of masses and with properties that 
dependent strongly on the global parameters of temperature, density, proton 
fraction and, in particular, also on the strength of the symmetry term 
under these conditions (Fig.~\ref{fig01}).

\section{Statistical description of isoscaling}
 
Isotopic scaling, also termed isoscaling, has been shown to be a
phenomenon common to many different types of heavy ion reactions
\cite{botv02,tsang01,soul03,fried04}.
It is observed by comparing product yields from otherwise identical
reactions with isotopically
different projectiles or targets, and it is constituted by an
exponential dependence of the yield ratios $R_{21}(N,Z)$ measured
for the two reactions
on the neutron number $N$ and proton number $Z$ of the produced fragments.
The scaling expression
\begin{equation}
R_{21}(N,Z) = Y_2(N,Z)/Y_1(N,Z) = C \cdot exp(\alpha \cdot N + \beta \cdot Z)
\label{eq:scalab}
\end{equation}
describes rather well the measured ratios for a wide range of
complex particles and light fragments \cite{tsang01a}.

In the grand-canonical approximation,
assuming that the temperature $T$ is about the same,
the scaling parameters $\alpha$ and $\beta$ are proportional
to the differences of the neutron and proton chemical potentials for
the two systems, $\alpha = \Delta \mu_{\rm n}/T$ and $\beta = \Delta \mu_{\rm p}/T$.
Of particular interest is the proportionality of the scaling parameters with
the symmetry energy term.
It has been obtained from the statistical interpretation of isoscaling
within the SMM \cite{botv02} and Expanding-Emitting-Source Model
\cite{tsang01a} and confirmed by an analysis of reaction dynamics
\cite{ono03}. The relation is
\begin{equation} \label{eq:dmunu}
\Delta \mu_{\rm n} = \mu_{\rm n}^1 - \mu_{\rm n}^2 \approx -4\gamma
(\frac{Z_{1}^2}{A_{1}^2}-\frac{Z_{2}^2}{A_{2}^2}) = -4\gamma \Delta (Z^2/A^2)
\end{equation}
where $Z_{1}$,$A_{1}$ and $Z_{2}$,$A_{2}$ are the charges and mass
numbers of the two systems. The difference of the chemical potential
depends essentially only on the coefficient $\gamma$ of the symmetry term
and on the isotopic compositions.

There is a simple physical 
explanation within the SMM why isoscaling should appear
in finite systems.
Charge distributions of fragments with fixed mass numbers $A$, as well 
as mass distributions for fixed $Z$, are approximately Gaussian with 
average values and variances which are connected with the temperature,
the symmetry coefficient, and other parameters. The mean values depend 
on the total mass and charge of the systems, e.g. via the chemical 
potentials in the grand canonical approximation, while 
the variances depend mainly on the physical conditions reached,
the temperature, the density and possibly other variables. For example, 
the charge variance $\sigma_Z\approx \sqrt(AT/8\gamma)$ obtained for
fragments with a given mass number $A$ in Ref. \cite{botv85} 
is only a function of the temperature and of the symmetry term coefficient
since the Coulomb contribution is very small.

The above relation opens the possibility for an experimental program 
for studying the evolution of the role of the symmetry term for the 
fragment formation as a function of the conditions of the experiment. 
Besides the isoscaling coefficient, the required inputs are the temperatures
at breakup and the difference of the neutron-to-proton ratios $N/Z$ of the
two systems. An encouraging result was obtained when this method was first 
applied to the data of light-ion (p, d, $\alpha$) induced reactions at 
relativistic energies of up to 15 GeV \cite{botv02}. The obtained value 
had about standard magnitude $\gamma \approx 23$~MeV. This is not unexpected in this
case because the data were inclusive. These reactions proceed mainly through 
heavy-residue formation, and the mean multiplicities of intermediate-mass 
fragments are correspondingly small \cite{beaulieu}. 

\section{$^{12}$C on $^{112,124}$Sn with INDRA@GSI}

Multifragmentation becomes a dominant channel in reactions of heavier 
projectiles or targets at relativistic energies. For 
the $^{12}$C on $^{112,124}$Sn reactions, studied with the INDRA multidetector
\cite{pouthas} in experiments performed at the GSI, maximum fragment 
production occurs at central impact parameters, according to the 
systematics \cite{schuett96}. The measurements were performed 
with enriched targets of $^{112}$Sn (98.9\%) and $^{124}$Sn (99.9\%). Reaction 
products with $Z \leq 5$ were detected and isotopically identified with 
high resolution by using the calibration telescopes of INDRA which are 
positioned at polar 
angles $45^{\circ} \leq \theta_{\rm lab} \leq 176^{\circ}$ \cite{lef05}.

\begin{figure}[htb]
\centering
\includegraphics[height=8.0cm]{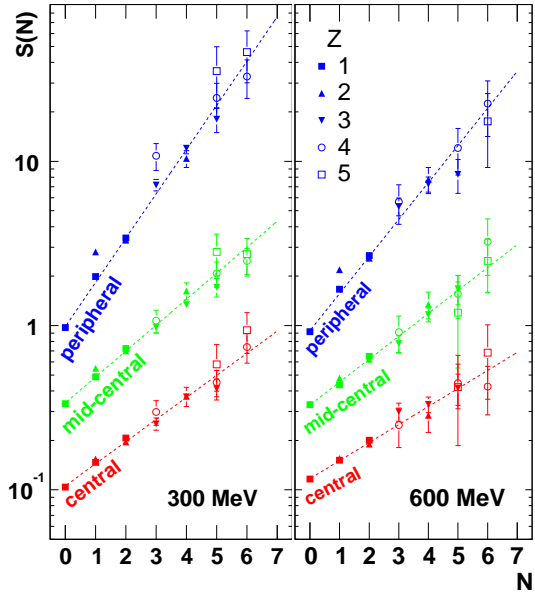}
\caption{
Scaled isotopic ratios $S(N)$ for $^{12}$C + $^{112,124}$Sn at $E/A$ = 300~MeV
(left panel) and 600~MeV (right panel) for three intervals of reduced
impact parameter with
''central'' indicating $b/b_{\rm max} \leq 0.4$ and with offset factors of
multiples of three.
The dashed lines are the results of exponential fits according
to Eq.~(\protect\ref{eq:scalab}). Only statistical errors are displayed
(from \protect\cite{lef05}).
}
\label{fig02} 
\end{figure}

The ratios of the energy-integrated fragment yields measured for the
two reactions and integrated over energy and angle
obey the laws of isoscaling.
This is illustrated in Fig.~\ref{fig02} for both incident energies and
for three impact parameter bins, selected on the basis of the charged-particle 
multiplicity measured with the full detector. The fitted dependence on
$Z$ has been used to scale the ratios so as to obtain a single branch
for each data set. The resulting slopes of the scaled ratios as a
function of the neutron number $N$ change considerably with impact parameter.
The obtained fit parameters extend from $\alpha$ = 0.61 to values as low as
$\alpha$ = 0.25 for the most central event group at 600 MeV per nucleon.

\begin{figure}[htb]
\centering
\begin{minipage}[c]{.49\textwidth}
   \centerline{\includegraphics[height=8.0cm]{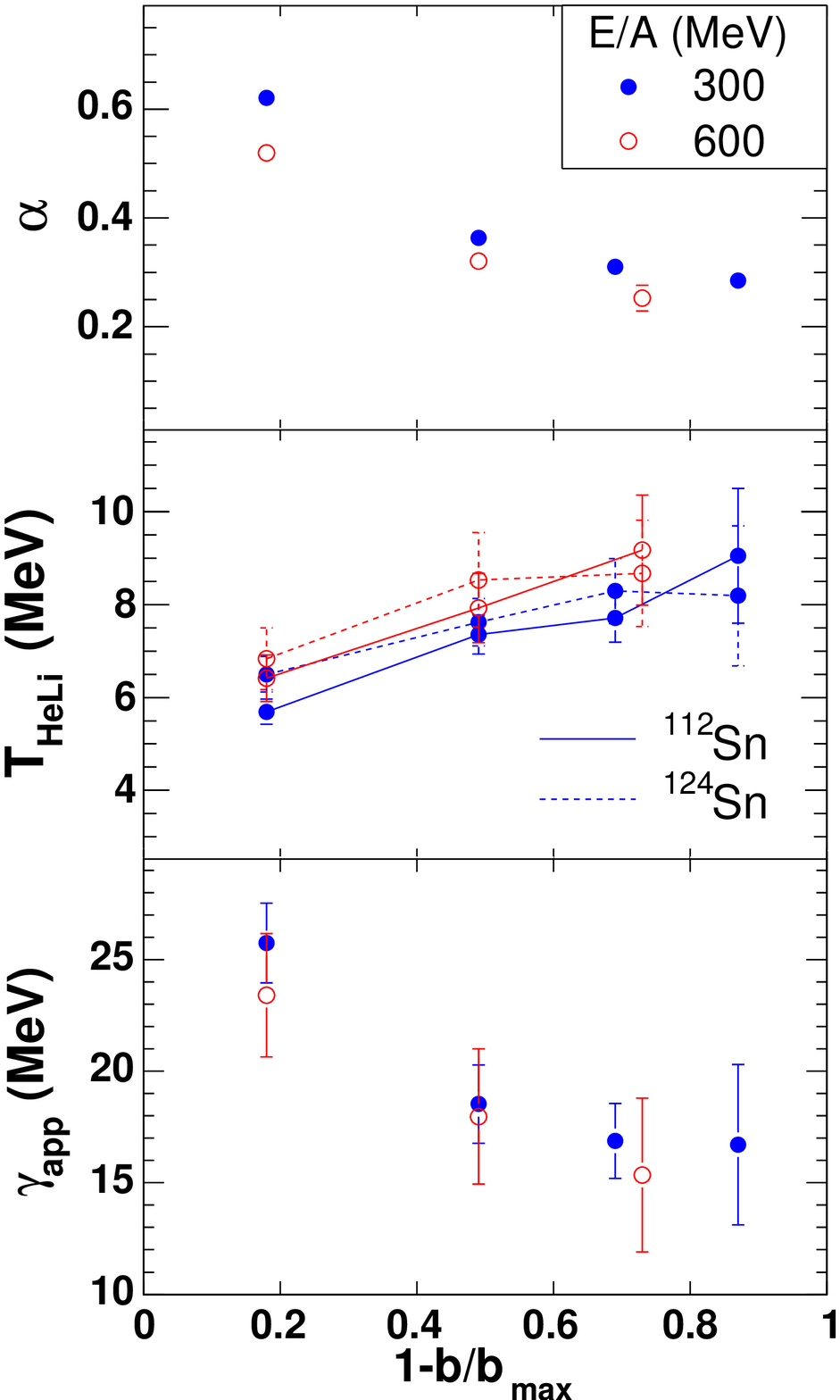}}
\end{minipage}
\begin{minipage}[c]{.49\textwidth}
   \centerline{\includegraphics[height=8.0cm]{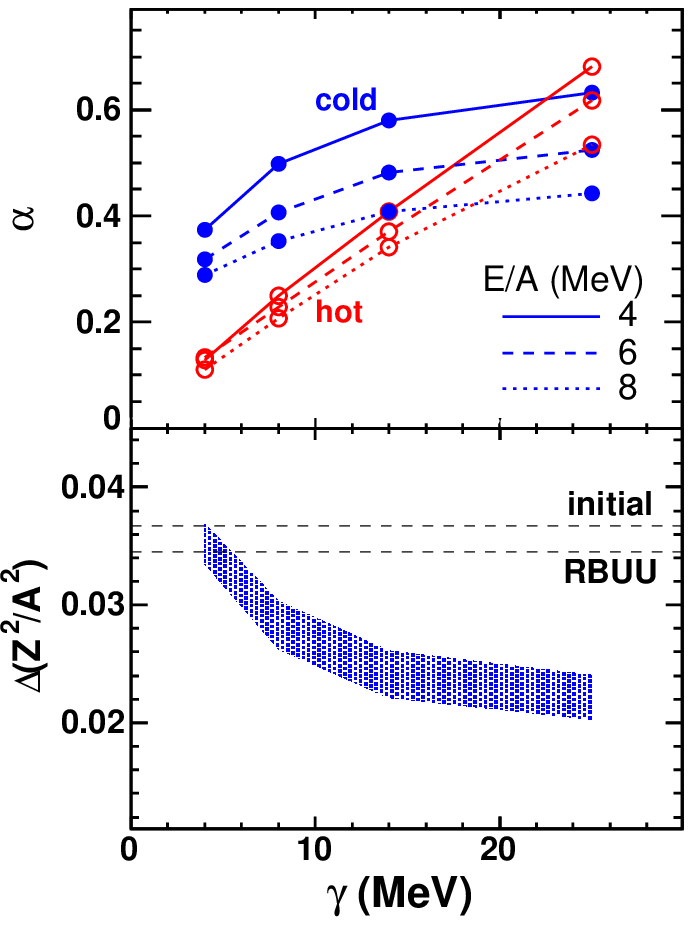}}
\end{minipage}

\caption{The left panels show the 
isoscaling coefficient $\alpha$ (top), the double-isotope temperatures $T_{\rm HeLi}$
(middle) and the resulting $\gamma_{\rm app}$ (bottom) for the $^{12}$C on $^{112,124}$Sn
reactions at $E/A$ = 300~MeV (full symbols) and 600~MeV (open symbols) 
as a function of the centrality parameter $1-b/b_{\rm max}$.
\newline 
The top right panel shows the isoscaling coefficient $\alpha$ for hot 
(open circles) and cold fragments (dots) as a function of the symmetry term 
coefficient $\gamma$ as predicted by the Markov-chain calculations for $^{112,124}$Sn.
The shaded area in the bottom right panel shows the region in the 
$\Delta (Z^2/A^2)$-versus-$\gamma$ plane that is consistent with the measured value 
$\alpha =0.29$ for central collisions
and with the Markov-chain predictions for cold fragments.
The dashed lines indicate the $\Delta (Z^2/A^2) = 0.0367$ of $^{112,124}$Sn and the 
RBUU prediction
(from \protect\cite{lef05}).
}
\label{fig03} 
\end{figure}

With temperature estimates obtained from the yields of $^{3,4}$He and $^{6,7}$Li
isotopes, the values for an apparent symmetry term $\gamma_{\rm app}$,
i.e. before sequential decay corrections,
were obtained according to Eq.~\ref{eq:dmunu}. For peripheral collisions,
they are close to the normal-density value $\gamma \approx$ 23 MeV
but drop to lower values at the more central impact parameters 
(Fig.~\ref{fig03}, bottom left).
The effects of sequential decay for the symmetry term were calculated within
the microcanonical Markov-chain version of the Statistical Multifragmentation
Model \cite{botv01} for excitation energies of 4, 6, and 8 MeV per nucleon
and for 4 MeV $\leq \gamma \leq$ 25 MeV.
The isoscaling coefficient $\alpha$ was
determined from the calculated fragment yields before (hot fragments) and
after (cold fragments) the sequential decay part of the code for which a
standard value $\gamma$ = 25 MeV was used.

The hot fragments exhibit the linear relation of $\alpha$ with $\gamma$ as expected
(Fig.~\ref{fig03}, top right).
For $\gamma$ smaller than 25 MeV, the sequential decay causes a narrowing of the 
initially broad isotope distributions which leads to an increase of the
isoscaling coefficients $\alpha$ for cold fragments.
The variation of $\alpha$ with $\gamma$ is thus considerably reduced 
with the effect that the value $\alpha < 0.3$ measured for the 
most central bins can only be reproduced with input values $\gamma \leq$ 10 MeV. 
This is illustrated in Fig.~\ref{fig03}, bottom right,
which also shows the effect of possible variations of the isotopic composition
of the two systems at breakup. Transport models predict that this difference
should not change by more than a few percent \cite{gait04}.

\section{Projectile fragmentation with ALADIN}

In a very recent experiment with the ALADIN spectrometer, the possibility 
of using secondary beams for reaction studies at relativistic energies 
has been explored \cite{sfi05}. 
Beams of $^{107}$Sn, $^{124}$Sn, $^{124}$La, and $^{197}$Au were
used to investigate the mass and isospin dependence of projectile 
fragmentation at 600 MeV per nucleon. The neutron-poor radioactive 
projectiles $^{107}$Sn and $^{124}$La were produced at the Fragment Separator FRS
by fragmentation of a primary beam of $^{142}$Nd and delivered to the ALADIN 
experiment. Natural Sn targets with areal density 500 mg/cm$^2$ were used in 
order to fully cover the rise and fall regimes of multifragmentation 
\cite{schuett96}. 
A cross sectional view of the experimental setup and a short description are
given in the contribution of De Napoli et al. to this workshop \cite{marzio}. 

\begin{figure}[htb]
\centering

\includegraphics[height=6.0cm]{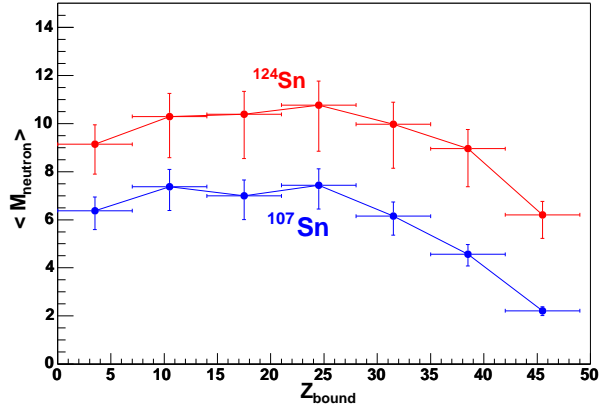}

\caption{
Neutron multiplicities of the spectator source for $^{107,124}$Sn projectile 
fragmentation as a function of $Z_{\rm bound}$ (preliminary result).
}
\label{fig05} 
\end{figure}

The mass resolution obtained for projectile fragments entering into the 
acceptance of the spectrometer is about 1.5\% for fragments with $Z\geq 6$
and increases up to about 3\% for the lightest fragments. The error in the
mass determination has about equal contributions from the measurement 
of the magnetic rigidity ($\Delta R/R\approx1\%$) and from the time-of-flight 
measurement ($\Delta t\approx 150$~ps, increasing up to 250 ps for light fragments). 
Masses are 
individually resolved for fragments with atomic number $Z \leq 10$. The elements
are individually resolved over the full range of atomic numbers up to the
projectile $Z$ with a resolution of $\Delta Z \leq 0.2$ obtained with the
TP-MUSIC IV detector \cite{sfi05,marzio}. 

Results of an analysis of the $Z$ fluctuations of the largest fragment 
within a partition are presented in \cite{marzio}. The preliminary analysis
of isotope ratios obtained for the studied reactions shows that isoscaling 
is observed. The decrease of the isoscaling parameters with increasing 
centrality is confirmed and extended over the larger fragmentation regime 
covered in the collisions with the Sn target. If normalized to the larger 
difference in $N/Z$ that is achieved by using radioactive beams 
the new data are found to be consistent with 
the results obtained for $^{12}$C on $^{112,124}$Sn. 

There is justified hope 
that the detection of projectile fragments without threshold 
with the ALADIN spectrometer will eventually permit the reconstruction
of the spectator system at breakup. Fragments with $Z \geq 3$ exhibit well 
defined rapidity distributions centered near projectile rapidity 
\cite{schuett96}. It will be necessary, however, to also identify the 
spectator sources of hydrogen and helium ions and of neutrons. 

Neutrons emitted in directions close to $\theta_{{\rm lab}} = 0^{\circ}$,
are detected with the Large-Area Neutron Detector (LAND) which covers 
about one half of the solid angle required for neutrons from the spectator 
decay. A preliminary 
analysis of the invariant multiplicity distributions of neutrons
has led to the identification of the spectator sources of neutrons. They are 
characterized by temperatures up to about 4 MeV possibly caused by 
large contributions from evaporation. Reconstructed multiplicities for 
$^{107,124}$Sn projectile fragmentation as a function of $Z_{\rm bound}$ are shown
in Fig.~\ref{fig05}. On average about 3 to 4 more neutrons are observed for
the more neutron rich spectator system. 

\section{Conclusion}

The present study of the role of the symmetry energy in spectator decay is 
motivated by the importance of this quantity for our understanding of 
astrophysical phenomena like supernovae and neutron stars. The similarity of 
the thermodynamic conditions reached in nuclear multifragmentation and in
supernova explosions gives experimental access to properties of hot 
supernova matter and its nuclear composition. 
The observed reduction of the symmetry term coefficient needed to describe
isotopic distributions is consistent with a global 
classification of multifragmentation as a low-density phenomenon. 
More specifically, it may indicate that small and intermediate-mass nuclei
are produced which are surrounded by other fragments and by a nucleon gas. The
structure of these highly excited fragments, at the chemical freeze-out
point, may be considerably different from that of stable nuclei.
The combined effects of the interaction with the neighbours and of their
own expanded structure provide sufficient reason for a strong reduction
of the symmetry term.

In order to draw firm conclusions the evolution of the spectator systems prior to breakup will have to be followed in more detail. The new experiments performed with the ALADIN spectrometer and with secondary beams offer the possibility to reconstruct the neutron-to-proton ratios of the systems at breakup. A spectator source of neutrons has been identified by analyzing the coincident data measured with LAND. The secondary decay of excited fragments has the effect of reducing the widths of the isotope distributions and of narrowing their overall separation in $N/Z$. Model calculations indicate that a reduction of the symmetry term will be partly masked by the sequential decay, with the effect that 
the true reduction at breakup may be considerably larger than what is observed for the final fragments.  

{\it The authors would like to thank T. Gaitanos for providing the results of
RBUU calculations for the studied systems.
C.Sf. acknowledges the receipt of an
Alexander-von-Humboldt fellowship.
This work was supported by the European Community under
contracts No. ERBFMGECT950083 and RII3-CT-2004-506078 
and by the Polish Scientific Research 
Committee under contract No. 2P03B11023.}



%
\end{document}